\documentclass[preprint,a4paper,11pt]{article}
\usepackage{jheppub}
\usepackage{hyperref}
\usepackage{lineno}

\title{\boldmath A minimal inverse seesaw model with $S_4$ flavour symmetry}

\author{Bikash Thapa,}
\author{Sunita Barman,}
\author{Sompriti Bora, and}
\author{Ng. K. Francis}
\affiliation{Department of Physics, Tezpur University, Tezpur - 784028, India}

\emailAdd{bikash2@tezu.ernet.in}
\emailAdd{sunitabarman282@gmail.com}
\emailAdd{sompriti.bora@gmail.com}
\emailAdd{francis@tezu.ernet.in}

\abstract{We construct an $S_4$ flavour symmetric minimal inverse seesaw model where the standard model is extended by adding two right-handed and two standard model gauge singlets neutrinos to explain the origin of tiny neutrino masses. The resulting model describes the lepton mass spectra and flavour mixing quite well for the case of the normal hierarchy of neutrino masses. The prediction of the model on the Dirac CP-violating phase is centered around $370.087^\circ$. Furthermore, using the allowed region for the model parameters, we have calculated the value of the effective Majorana neutrino mass, $\lvert\langle m_{ee}\rangle\rvert$, which characterizes neutrinoless double beta decay.}


\begin{document}
\maketitle
\flushbottom

\section{Introduction}
\label{sec:intro}
Since the discovery of neutrino oscillation in various experiments \cite{PhysRevLett.107.181802, PhysRevLett.107.041801, PhysRevLett.108.171803, PhysRevLett.108.191802}, it is established that neutrinos are massive and their flavours are mixed. On the other hand, neutrinos are massless in the Standard Model (SM), this points towards the existence of physics beyond the SM. There exist various theories that can explain the origin of tiny neutrino masses such as the seesaw mechanism \cite{mohapatra1980neutrino, mohapatra1981neutrino, minkowski1977mu, yanagida1980horizontal, mohapatra1986neutrino}, radiative seesaw mechanism \cite{ma2006verifiable}, and models based on extra dimensions \cite{ma2000light, arkani2001neutrino}. One such framework is the inverse seesaw (ISS) mechanism, where the SM is extended by introducing SM gauge singlets and right-handed (RH) neutrinos \cite{gonzalez1990isosinglet, deppisch2005enhanced}. In contrast to the canonical seesaw mechanism, in the ISS mechanism, besides the heaviness of RH neutrinos, a small lepton-number violating mass parameter $\mu$ causes the suppression of the neutrino mass allowing lighter RH neutrinos (TeV-scale) and $\mathcal{O}$(1) Yukawa coupling. The gauge-invariant Lagrangian of the extension of the SM can be written as
\begin{equation}
\label{eq:1.1}
	-\mathcal{L}_\nu = Y_\nu \bar{l}_L \tilde{H} N_R + M_R \bar{\left(N_R\right)^c} \left(S_L\right)^c + \frac{1}{2} \mu \bar{S_L} \left(S_L\right)^c + h.c.,
\end{equation}

where $l_L$ is the left-handed doublet, $H$ is the Higgs doublet,$\tilde{H} = i \sigma_2H^*$ with $\sigma_2$ being the $2^{nd}$ Pauli matrix, $N_R$ are the right-handed neutrino singlets and $S_L$ are the SM gauge singlets. After the Higgs doublet, $H$ acquires vacuum expectation value ($vev$), i.e., $\langle H \rangle = (0, v/\sqrt{2})^T$ with $v \approx 246$ GeV and breaks the gauge symmetry, the neutrino mass matrix may be written as 

\begin{equation}
\label{eq:1.2}
	M_\nu =
		\begin{pmatrix}
			0     & m_D   & 0   \\
			m_D^T & 0     & M_R \\
			0     & M_R^T & \mu
		\end{pmatrix},
\end{equation}

where $m_D = Y_\nu v/\sqrt{2}$ is the Dirac mass matrix, $M_R$ is a complex matrix and $\mu$ is a complex, symmetric matrix. With $\mu << m_D << M_R$, diagonalisation of eq. \ref{eq:1.1} lead to

\begin{equation}
\label{eq:1.3}
	m_\nu = m_D \left(M_R^T\right)^{-1} \mu \left(M_R\right)^{-1} m_D^T .
\end{equation}

In the ISS model, with the $\mathcal{O}(m_D)\sim 10^2$ GeV, the mass of light neutrinos $\mathcal{O}(m_\nu) \sim 0.1$ eV may be suppressed by the smallness of $\mu$ with $\mathcal{O}(\mu)\sim 10^3$ eV as well as the heaviness of right-handed neutrino masses $\mathcal{O}(M_R) \sim 10^4$ GeV. The mass scale of the heavy neutrinos is slightly lower than the canonical seesaw model making it potentially testable at future colliders. 

Another aspect of the flavour structure of the SM is the observed fermion mixing. For the lepton sector, experimental evidence shows two large and a small mixing angles, however, the origin of such mixing patterns is still unanswered. The answer to such a problem can be given by introducing non-Abelian discrete flavour symmetries into the Lagrangian of the model \cite{morisi2013neutrino, king2013neutrino, king2017unified}. Various models based on $A_4$ \cite{brahmachari20084, altarelli2010discrete, lei2020minimally, barman2023nonzero, vien2015neutrino, vien2020cobimaximal, vien2022b, vien20224}, $S_4$ \cite{mohapatra2004high, hagedorn2006s4, cai2006s, bazzocchi2009s, yang2011minimal, bazzocchi2013neutrino, penedo2017neutrino, thapa2021resonant, dong20113, vien2014new, vien2016lepton, vien2019fermion, vien2021multiscalar, vien2022renormalizable}, $A_5$ \cite{ding2012golden, ballett2015mixing}, etc. have been proposed over the years to explain the observed lepton flavour mixing pattern. In models based on non-Abelian discrete symmetries, the discrete symmetry which is exact at a high-energy scale breaks down distinctly leaving residual symmetry in the charged-lepton and neutrino sectors at low-energy scales. This breaking pattern is governed by the $vev$ of the scalar field known as flavons (singlets under SM gauge symmetry) and eventually determines the lepton flavour mixing pattern.

In this paper, we study the ISS model with $S_4$ flavour symmetry and examine how well the model describes neutrino masses, mixing, and CP violation. We work in the framework with minimal ISS(2, 2) which is the minimal possible form of ISS mechanism that can account for the neutrino mass spectra \cite{abada2014looking}. The resulting neutrino mass matrix is tested against the neutrino experimental data using chi-squared analysis. We further explore the implications of the model for neutrinoless double beta decay.  

The rest of the article is structured as follows. In section \ref{sec:2}, we construct the $S_4$ flavour symmetric inverse seesaw model with two right-handed and two SM gauge singlet neutrinos. Section \ref{sec:3} includes the numerical analysis and the results of the model presented in section \ref{sec:2}. We investigate the viability of the model to explain the latest data from neutrino oscillation experiments using chi-squared analysis. Further, we define the allowed region of the parameters of the model corresponding to $\chi^2 \leq 30$ values. This section also includes the results on neutrinoless double beta decay predicted by the model and we finally summarise our conclusions in section \ref{sec:4}.	 

\section{The model}
\label{sec:2}

We consider the extension of the SM by including additional $S_4$ flavour symmetry. It is further augmented with $Z_3 \times Z_4$ group to achieve the desired structures for the mass matrices. The fermion sector of the model includes the addition of two right-handed neutrinos and two SM gauge singlet fermions to the SM fermion content, resulting in ISS(2, 2) framework. In the scalar sector, we have one $SU(2)_L$ Higgs doublet $H$, and $SU(2)_L$ singlet flavons $\phi_c$, $\varphi_c$, $\phi_\nu$, $\chi$, $\psi$. Various fields of the model and their transformation properties under different symmetry groups are presented in Table \ref{tab:1}. The Yukawa Lagrangian which is invariant under the flavour symmetry, is of the form

\begin{align}
\label{eq:2.1}
	-\mathcal{L}  \supset &\frac{\alpha_1}{\Lambda}\bar{l}_L H \varphi_c e_R + \frac{\alpha_2}{\Lambda}\bar{l}_L H \varphi_c (\mu_R, \tau_R) + \frac{\alpha_3}{\Lambda}\bar{l}_L H \phi_c (\mu_R, \tau_R)\\ \nonumber
	             &+ \frac{\beta_1}{\Lambda} \bar{l}_L \tilde{H} \varphi_\nu N_R + \gamma_1 \bar{N}_R \xi S_1  + \gamma_2 \bar{N}_R \xi S_2
	             + \lambda_1 S_1 S_1 \psi + \lambda_2 S_2 S_2 \psi + h.c., 
\end{align}

where  $\alpha_1$, $\alpha_2$, $\alpha_3$, $\beta_1$, $\gamma_1$, $\gamma_2$, $\lambda_1$, and $\lambda_2$ are the Yukawa coupling constant.

The $vev$ of the flavons in the charged-lepton sector are $\langle \varphi_c \rangle = (v_{\varphi_c}, 0, 0)$, $\langle \phi_c \rangle = (v_{\phi_c}, 0, 0)$ \cite{zhao2011realizing}. The charged-lepton mass matrix obtained after flavour and electroweak symmetry breaking is of the form,

\begin{equation}
\label{eq:2.2}
	m_l = \frac{v}{\Lambda}
	      \begin{pmatrix}
	      	\alpha_1 v_{\varphi_l} & 0 & 0\\
	      	0 & \alpha_2 v_{\varphi_l} + \alpha_3 v_{\phi_l} & 0\\
	      	0 & 0 & \alpha_2 v_{\varphi_l} - \alpha_3 v_{\phi_l}
	      \end{pmatrix}.
\end{equation} 

\renewcommand{\arraystretch}{1.2}
\begin{table}[t]
	\centering
	\begin{tabular}{ c | c c c c c c c c c c c c }
	\hline
	Field & $l_L$ & $e_R$ & ($\mu_R$, $\tau_R$) & $H$ & $N_R$ & $S_1$ & $S_2$ & $\varphi_c$ & $\phi_c$ & $\varphi_\nu$ & $\xi$ & $\psi$\\
	\hline
	S$_4$ & 3$_1$ & 1$_1$ & 2 & 1$_1$ & $2$ & 1$_1$ & 1$_2$ & 3$_1$ & 3$_2$ & 3$_1$ & 2 & 1$_1$\\
		
	Z$_3$ & 1 & $\omega^2$ & $\omega^2$ & 1 & 1 & 1& 1 & $\omega$ & $\omega$ & 1 & 1 & 1\\

	Z$_4$ & $i$ & 1 & 1 & 1 & $-i$ & $i$ & $i$ & $-i$ & $-i$ & 1 & 1 & -1\\
	\hline
	\end{tabular}
	\caption{Field content of the model and their charge assignment under S$_4  \times $ Z$_3 \times $Z$_4$.}
	\label{tab:1}
\end{table}

The hierarchy among the masses of the charged-lepton can be explained using the Froggatt-Nielsen mechanism and we have assumed the approach presented in \cite{zhao2011realizing}. For the neutrino sector, we assume that the flavons develop $vev$ in a region of the scalar potential's parameter space where

\begin{equation}
\label{eq:2.3}
	\langle \varphi_\nu \rangle = (v_{\varphi_{\nu_1}}, v_{\varphi_{\nu_2}}, v_{\varphi_{\nu_3}}),~~~\langle \xi \rangle = (v_\xi, v_\xi),~~~\langle \psi \rangle = v_\psi.
\end{equation}

After electroweak gauge and flavour symmetry breaking, we get the following matrices for the mass term and couplings

\begin{equation}
\label{eq:2.4}
	m_D = 
	\begin{pmatrix}
		b & c \\
		a & b \\
		c & a
	\end{pmatrix},~~~	
	M_R =
	\begin{pmatrix}
		d & d\\
		d &-d
	\end{pmatrix},~~~	
	\mu =
	\begin{pmatrix}
		e & 0\\
		0 & e
	\end{pmatrix},
\end{equation}

where $a = \frac{\beta_1}{\Lambda} v v_{\varphi_{\nu_1}}$, $b = \frac{\beta_1}{\Lambda} v v_{\varphi_{\nu_2}}$, $c = \frac{\beta_1}{\Lambda} v v_{\varphi_{\nu_3}}$, $d = \gamma_1 v_\xi \simeq \gamma_2 v_\xi$, and $e = \lambda_1 v_\psi \simeq \lambda_2 v_\psi$.

Using the matrices of eq. \ref{eq:2.4} in the inverse seesaw formula (eq. \ref{eq:1.3}), the light neutrino mass matrix becomes

\begin{equation}
\label{eq:2.5}
	m_\nu = m_0 
	\begin{pmatrix}
		1+\beta^2 & \alpha+\beta & \beta(1+\alpha)\\
		\alpha+\beta & 1+\alpha^2 & \alpha(1+\beta)\\
		\beta(1+\alpha) & \alpha(1+\beta) & \alpha^2+\beta^2
	\end{pmatrix},
\end{equation}

where we have defined two complex dimensionless parameters $\alpha = a/b$, $\beta = c/b$ and the factor $m_0$ denotes the mass scale. The light neutrino mass matrix of eq. \ref{eq:2.5} is diagonalized by the Pontecorvo-Maki-Nakagawa-Sakata (PMNS) mixing matrix, $U_{\textrm{PMNS}}$

\begin{equation}
\label{eq:2.6}
	U_{\textrm{PMNS}}^\dagger m_\nu U_{\textrm{PMNS}}^* = \textrm{diag(}m_1, m_2, m_3 \textrm{)},
\end{equation}

with $m_1$, $m_2$, and $m_3$ being the mass eigenvalues. In standard PDG parametrization, the PMNS mixing matrix is given by

\begin{equation}
\label{eq:2.7}
	U_{\textrm{PMNS}} = 
	\begin{pmatrix}
	c_{12} c_{13} & s_{12} c_{13} & s_{13} e^{-i \delta_{CP}} \\
	-s_{12}c_{23}-c_{12}s_{23}s_{13}e^{i\delta_{CP}} & c_{12}c_{23}-s_{12}s_{23}s_{13}e^{i\delta_{CP}} & s_{23}c_{13} \\
	s_{12}c_{23}-c_{12}c_{23}s_{13}e^{i\delta_{CP}} & -c_{12}s_{23}-s_{12}c_{23}s_{13}e^{i\delta_{CP}} & c_{23}c_{13}
	\end{pmatrix}
	P_M,
\end{equation}
where, 

\begin{equation}
\label{eq:2.8}
	P_M = 
	\begin{pmatrix}
	e^{i\rho} & 0 & 0 \\
	0 & e^{i\sigma} & 0 \\
	0 & 0 & 1
	\end{pmatrix}.
\end{equation}

In the model under study, the lightest neutrino mass $m_1$($m_3$) is zero in the case of the Normal Hierarchy (Inverted Hierarchy) of neutrino masses. It is worth noting that for $m_1 = 0$ (NH), the Majorana CP phase $\rho$ is zero and in the case of IH with $m_3 = 0$ the phases can be redefined as $\sigma - \rho$ as $\sigma$. Thus, the model has a single Majorana CP phase and the phase matrix effectively becomes $P_M = diag(1, e^{i\sigma}, 1)$.

\section{Numerical analysis and results} 
\label{sec:3}
As discussed in the previous section, we have considered the extension of SM by including two RH neutrinos and two SM gauge singlets resulting in a framework known as the ISS(2,2) model. We have shown how $S_4$ flavour symmetry can be implemented in such a framework and we have obtained the light neutrino mass matrix as shown in eq. \ref{eq:2.5}. The neutrino mass matrix of eq. \ref{eq:2.5} contains four real parameters (Re($\alpha$), Im($\alpha$), Re($\beta$), Im($\beta$)) that effect the neutrino mixing matrix elements. In this section, we perform numerical analysis and test the model against the experimental data. We proceed by writing the neutrino oscillation parameters ($\theta_{12}$, $\theta_{23}$, $\theta_{13}$, $\Delta m_{21}^2$, $\Delta m_{31(32)}^2$, $\delta_{CP}$) in terms of the model parameters and scrutinize the ability of the model to explain the  neutrino experimental data \cite{esteban2020fate}. 

Using the light neutrino mass matrix we can define the hermitian matrix, $h_\nu = m_\nu m_\nu^\dagger$ such that 

\begin{equation}
\label{eq:3.1}
	h_\nu = m_0^2
			\begin{pmatrix}
				A   & B   & C \\
				B^* & D   & E \\
				C^* & E^* & F
			\end{pmatrix},
\end{equation}

with,
\begin{align*}
	A = & \lvert 1+\beta^2 \rvert^2 + \lvert \alpha+\beta \rvert^2 + \lvert \beta(1+\alpha) \rvert^2\\
	B = & (1+\beta^2)(\alpha+\beta)^* + (\alpha+\beta)(1+\alpha^2)^* + (\beta+\alpha\beta)(\alpha+\alpha\beta)^*\\
	C = & (1+\beta^2)(\beta+\alpha\beta)^* + (\alpha+\beta)(\alpha+\alpha\beta)^* + (\beta+\alpha\beta)(\alpha^2+\beta^2)^*\\
	D = & \lvert (\alpha+\beta) \rvert^2 + \lvert (1+\alpha^2) \rvert^2 + \lvert (\alpha+\alpha\beta) \rvert^2 \\
	E = & (\alpha+\beta)(\beta+\alpha\beta)^* + (1+\alpha^2)(\alpha+\alpha\beta)^* + (\alpha+\alpha\beta)(\alpha^2+\beta^2)^*\\
	F = & \lvert \beta(1+\alpha) \rvert^2 + \lvert (\alpha+\alpha\beta) \rvert^2 + \lvert (\alpha^2+\beta^2) \rvert^2.
\end{align*}

The analytical relations between the elements of the hermitian matrix $h_\nu$ and the three mixing angles as well as the Dirac CP-violating phase can be written as \cite{xing2010generic}

\renewcommand{\arraystretch}{1.5}
\begin{table}[t]
	\centering
	\begin{tabular}{ l c c }
	\hline
	Parameter                                           & Best-fit $\pm 1\sigma$          & $3\sigma$ range    \\
	\hline
	$\sin^2\theta_{12}$                                 & $0.304^{+0.013}_{-0.012}$       & 0.269 - 0.343      \\
	$\sin^2\theta_{23}$ (NH)                            & $0.573^{+0.018}_{-0.023}$       & 0.405 - 0.620      \\   
	$\sin^2\theta_{23}$ (IH)                            & $0.578^{+0.017}_{-0.021}$       & 0.410 - 0.623      \\
	$\sin^2\theta_{13}$ (NH)                            & $0.02220^{+0.00068}_{-0.00062}$ & 0.02034 - 0.02340  \\
	$\sin^2\theta_{13}$ (IH)                            & $0.02238^{+0.00064}_{-0.00062}$ & 0.02053 - 0.02434  \\
	$\frac{\Delta m_{21}^2}{10^{-5}\textrm{eV}^2}$      & $7.42^{+0.21}_{-0.20}$           & 6.82 - 8.04        \\
	$\frac{\lvert\Delta m_{3l}^2\rvert}{10^{-3}\textrm{eV}^2}$ (NH) & $2.515^{+0.028}_{-0.028}$        & 2.431 - 2.599      \\ 	
	$\frac{\lvert\Delta m_{3l}^2\rvert}{10^{-3}\textrm{eV}^2}$ (IH) & $2.498^{+0.028}_{-0.029}$        & 2.584 - 2.413      \\
	\hline
	\end{tabular}
	\caption{Latest experimental data on neutrino oscillation considered in our analysis \cite{esteban2020fate}.}
	\label{tab:2}
\end{table}

\begin{align}
\label{eq:3.2}
	&\tan \theta_{23} = \frac{Im(B)}{Im(C)}\nonumber \\
	&\tan 2\theta_{12}= \frac{2 N_{12}}{N_{22}-N_{11}} \nonumber \\
	&\tan \theta_{13} = \lvert Im(E)\rvert \cdot\frac{\sqrt{\{\left[Im(B)\right]^2 + \left[Im(C)\right]^2\}^2 + \{Re(B)Im(B) + Re(C)Im(C)\}^2}}{\sqrt{\{\left[Im(B)\right]^2 + \left[Im(C)\right]^2\} \{Re(B)Im(C) - Im(B)Re(C)\}^2}} \nonumber \\
	&\tan \delta_{CP} = -\frac{\left[Im(B)\right]^2 + \left[Im(C)\right]^2}{Re(B)Im(B)+Re(C)Im(C)} ,
\end{align}  
where the quantities $N_{11}$, $N_{12}$, and $N_{22}$ is expressed as
\begin{align}
\label{eq:3.3}
	N_{11} &= A - \frac{Re(B)Im(C)-Im(B)Re(C)}{Im(E)} \nonumber \\
	N_{12} &= \left[\frac{\left[Re(B)Im(C)-Im(B)Re(C)\right]^2}{\left[Im(B)\right]^2 + \left[Im(C)\right]^2} + \left[\frac{\{Re(B)Im(B)+Re(C)Im(C)\}^2}{\{\left[Im(B)\right]^2 + \left[Im(C)\right]^2\}^2} + 1\right]\left[Im(E)\right]^2\right]^{\frac{1}{2}} \nonumber \\
	N_{22} &= \frac{\left[Im(C)\right]^2 D + \left[Im(B)\right]^2 F - 2Im(B)Im(C)Re(E)}{\left[Im(B)\right]^2 + \left[Im(C)\right]^2}
\end{align}
It is clear that for a specific point in the four-dimensional parameter space of the model, there is a certain value of the experimental observables given by eq. (\ref{eq:3.2}). Consequently, any variation in the model parameters changes the value of the neutrino oscillation parameters resulting from the model. In order to test the model against the latest experiment data on neutrino mixing parameter, we define a $\chi^2$-function and perform a numerical simulation using a sampling package \texttt{MultiNest} \cite{feroz2009multinest}. The $\chi^2$-function used in our analysis has the following form
\begin{equation}
\label{eq:3.4}
	\chi^2 = \sum_i\left(\frac{P_i(p)-P_i^0}{\sigma_i}\right)^2,
\end{equation}
where $P_i(p)$ is the value of the observables predicted by the model at a point $p$ in the four-dimensional parameter space of the model, $P_i^0$ and $\sigma_i$ denotes the central value, and the corresponding 1$\sigma$ error of the $i^{th}$ experimental observable. The experimental values of the neutrino observables used in our analysis are summarised in Table \ref{tab:2}. In eq. \ref{eq:3.4}, we do not consider the Dirac CP-violating phase $\delta_{CP}$ as an input. The reason is the weak statistically preferred value of maximally violating CP phase from global experimental data. In order to carry out the test we treat the parameters of the model to be free and allow them to randomly vary in the following range

\begin{equation}
\label{eq:3.5}
	Re(\alpha),~ Im(\alpha),~ Re(\beta),~ Im(\beta) \in [-10,10].
\end{equation} 

Using eq. \ref{eq:3.2}, we obtain values for the three mixing angles and the Dirac CP violating phase. The best-fit values of the model parameter correspond to the minimum value of $\chi^2$. We found that the model gives a good description of the experimental data for NH of neutrino masses with $\chi^2_{min} \approx 0.24$, however, fails to describe the data for IH with $\chi_{min}^2 > 100$.
\begin{figure}[t]
	\begin{center}
		\includegraphics[width = \textwidth]{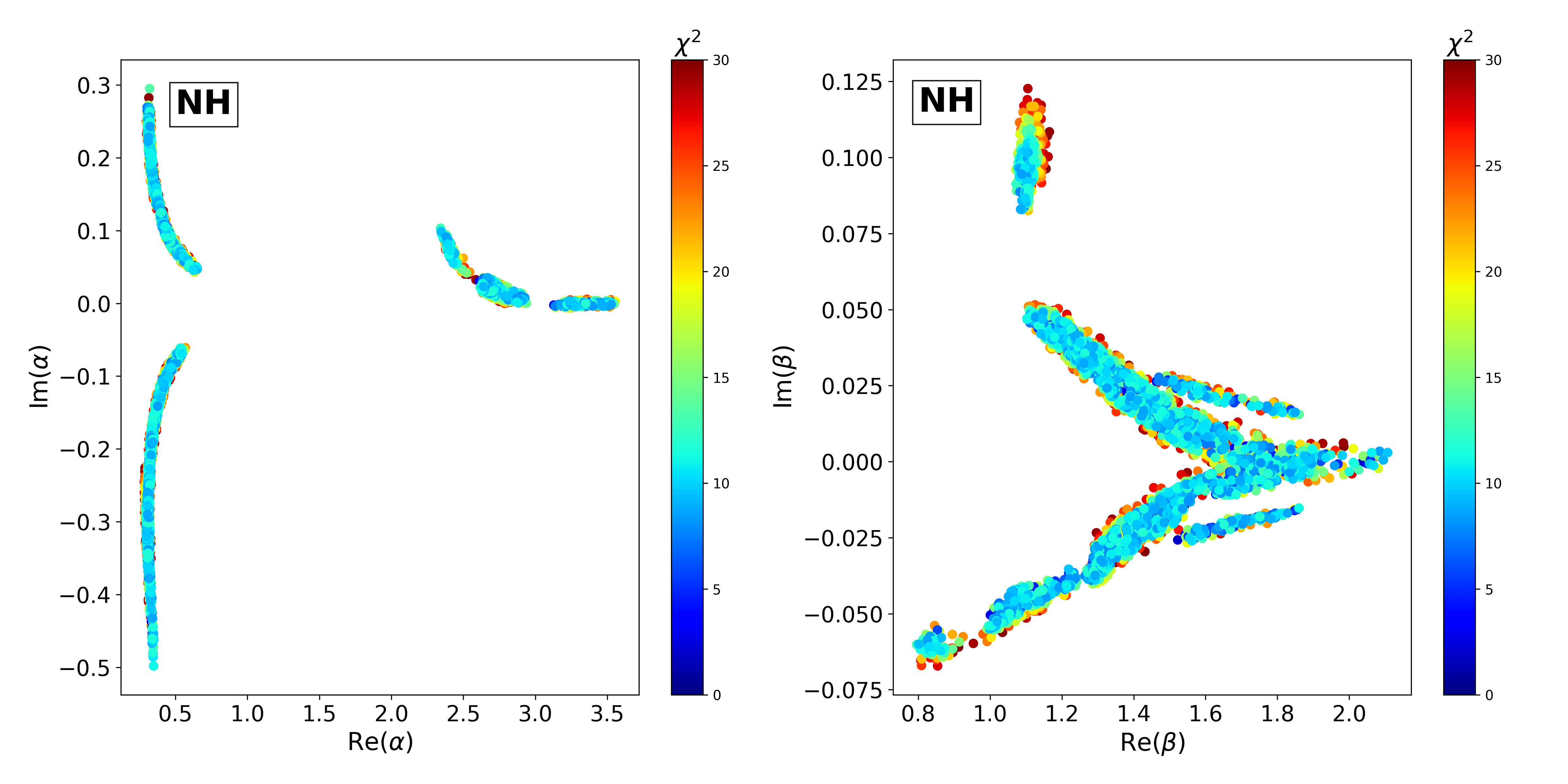}
		\caption{Allowed region for the model parameters $Re(\alpha)$, $Im(\alpha)$, $Re(\beta)$, and $Im(\beta)$.}
		\label{fig:1}
	\end{center}
\end{figure} 
The allowed region of the parameter space of the model is shown in Figure \ref{fig:1} with the colors indicating the range of values of $\chi^2$. Here, we have shown the values of the parameters of the model corresponding to $\chi^2 \leq 30$. The best-fit values of the model parameters obtained by minimizing the $\chi^2$-function are $Re(\alpha) = 0.314$, $Im(\alpha) = -0.255$, $Re(\beta) = 1.293$, and $Im(\beta) = 0.032$ in NH \footnote{In our analysis we have accepted only the points that satisfy $\chi^2 \leq 30$, hence, no further analysis is done for the case of IH.}.

\begin{figure}[t]
	\begin{center}
		\includegraphics[width = \textwidth]{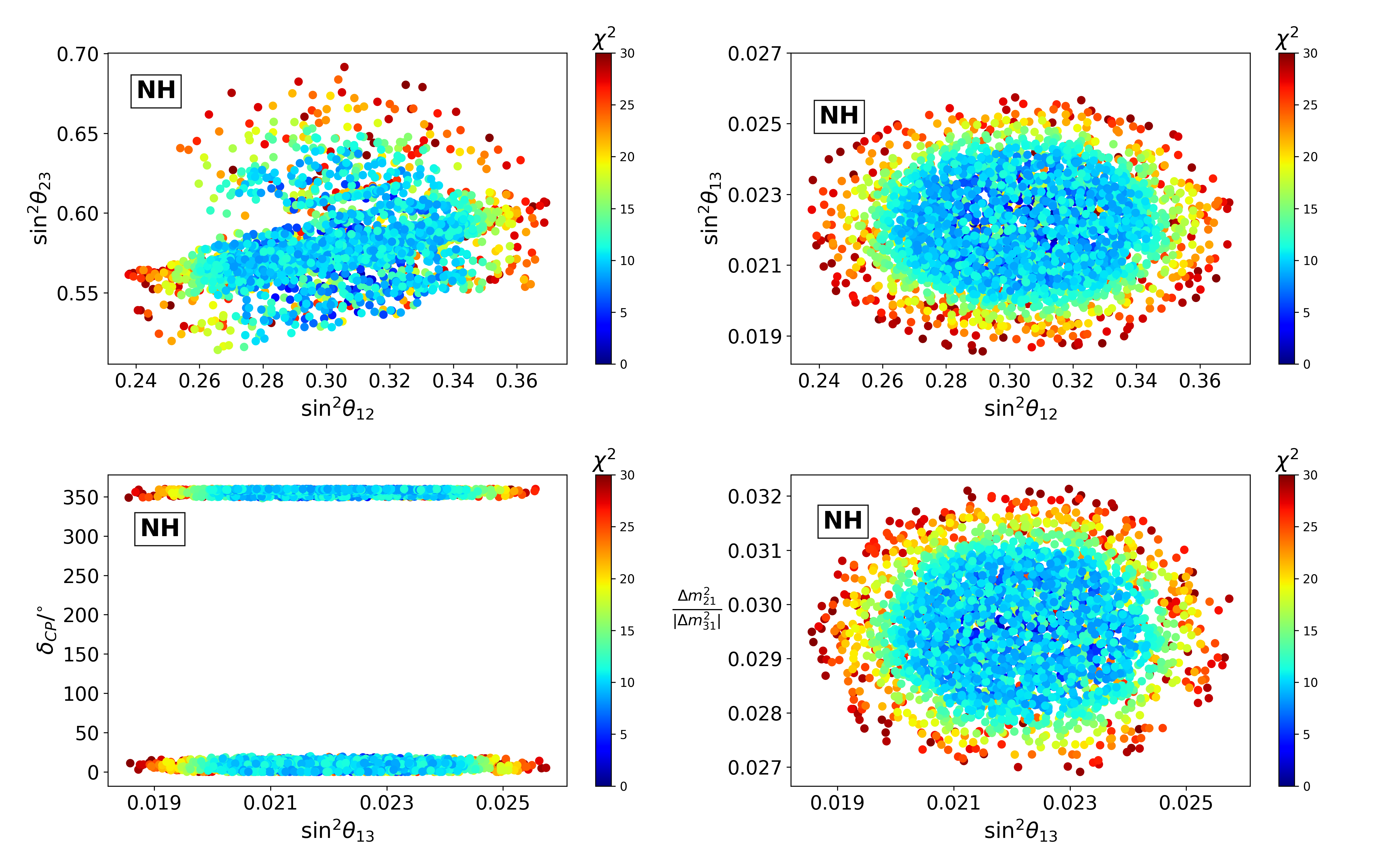}
		\caption{Correlation between the neutrino oscillation parameters with the color indicating the ranges of $\chi^2$ values.}
		\label{fig:2}
	\end{center}
\end{figure} 

\begin{table}[t]
	\centering
	\begin{tabular}{ c c c c c c }
	\hline
	$\sin^2\theta_{12}$ & $\sin^2\theta_{23}$ & $\sin^2\theta_{13}$ & $\delta_{CP}/^\circ$ & $\Delta m_{21}^2$ ($eV^2$) & $\Delta m_{31}^2$ ($eV^2$)\\
	\hline	
	0.303 & 0.575 & 0.0225 & 370.087 & $7.42 \times 10^{-5}$ & $2.510 \times 10^{-3}$\\
	\hline
	\end{tabular}
	\caption{The best-fit values for the neutrino oscillation parameters from $\chi^2$ analysis. }
	\label{tab:3}
\end{table}

Figure \ref{fig:2} shows how well the model describes the neutrino oscillation experimental data for the case of NH of neutrino masses. The color bar represents the value of $\chi^2$ ranging from ($1-30$). The best-fit values of the neutrino oscillation parameters obtained from the model are summarized in Table \ref{tab:3}. The best-fit values for the parameters $\sin^2 \theta_{12}$, $\sin^2 \theta_{23}$, $\sin^2 \theta_{13}$ and the two mass-squared differences lies well within the $1\sigma$ range of experimental values shown in Table \ref{tab:2}. The value of the Dirac CP-violating phase $\delta_{CP}$ corresponding to the $\chi^2_{min}$ value is $370.078^\circ$, which is within the $3\sigma$ range of neutrino oscillation data. Thus the model presented in the previous section provides a decent description of the recent experimental data and whose prediction on $\delta_{CP}$ may be tested in future precision experiments.

\begin{figure}[t]
	\begin{center}
		\includegraphics[width = 0.8\textwidth]{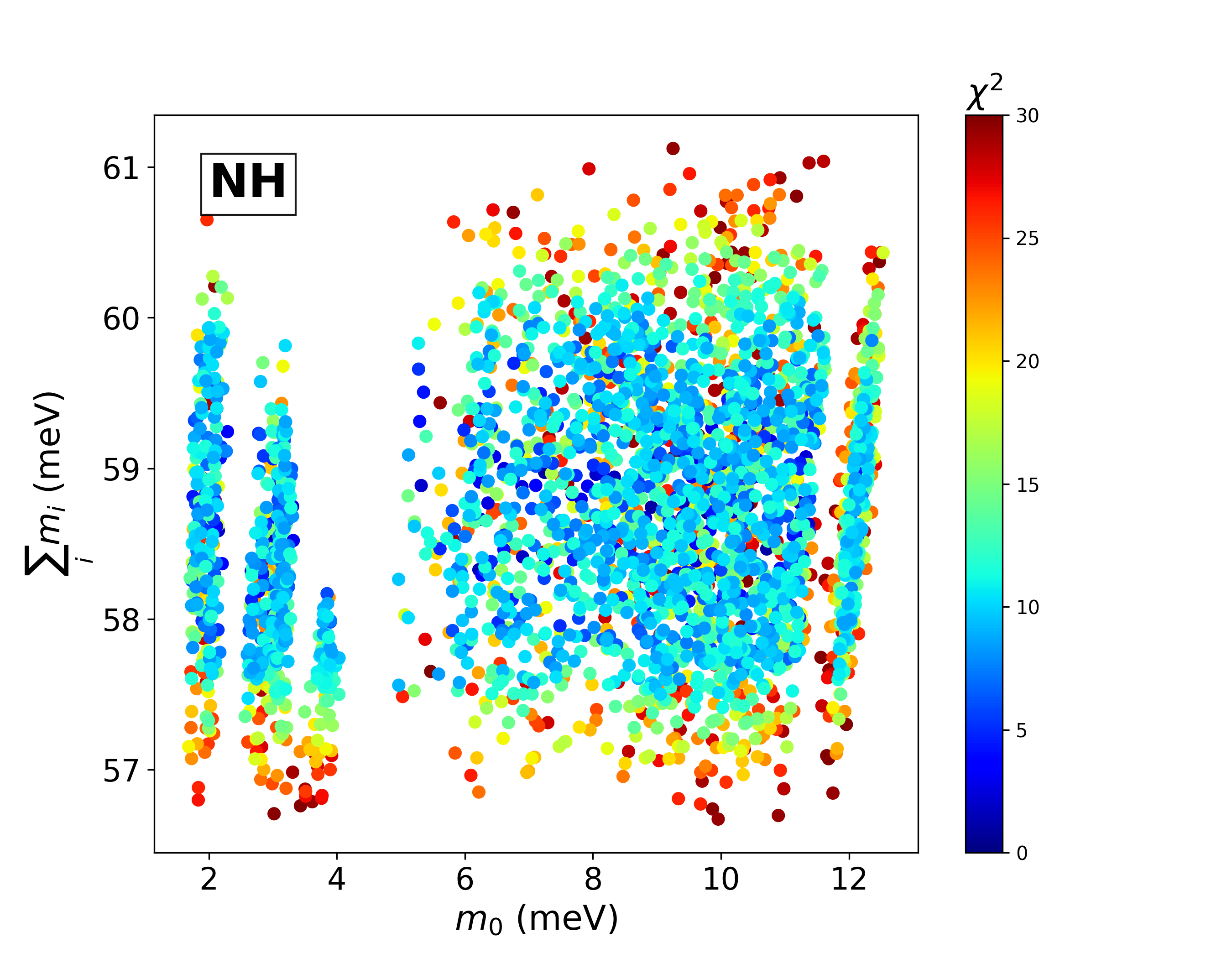}
		\caption{Range of values for the mass scale, $m_0$ and the sum of light neutrino masses, $\sum_i m_i$.}
		\label{fig:3}
	\end{center}
\end{figure}

In Figure \ref{fig:3} we present the sum of neutrino masses as a function of the mass scale, $m_0$ which effectively influences the absolute neutrino masses. There is a cosmological upper bound on the sum of the light neutrino masses, $\sum_i m_i < 0.12$ eV \cite{PhysRevD.96.123503, PhysRevD.98.123526, tanseri2022updated, aghanim2020planck} and our model shows a consistent value ranging from $56.67$ to $61.12$ meV. 

\begin{figure}[t]
	\begin{center}
		\includegraphics[width = 0.8\textwidth]{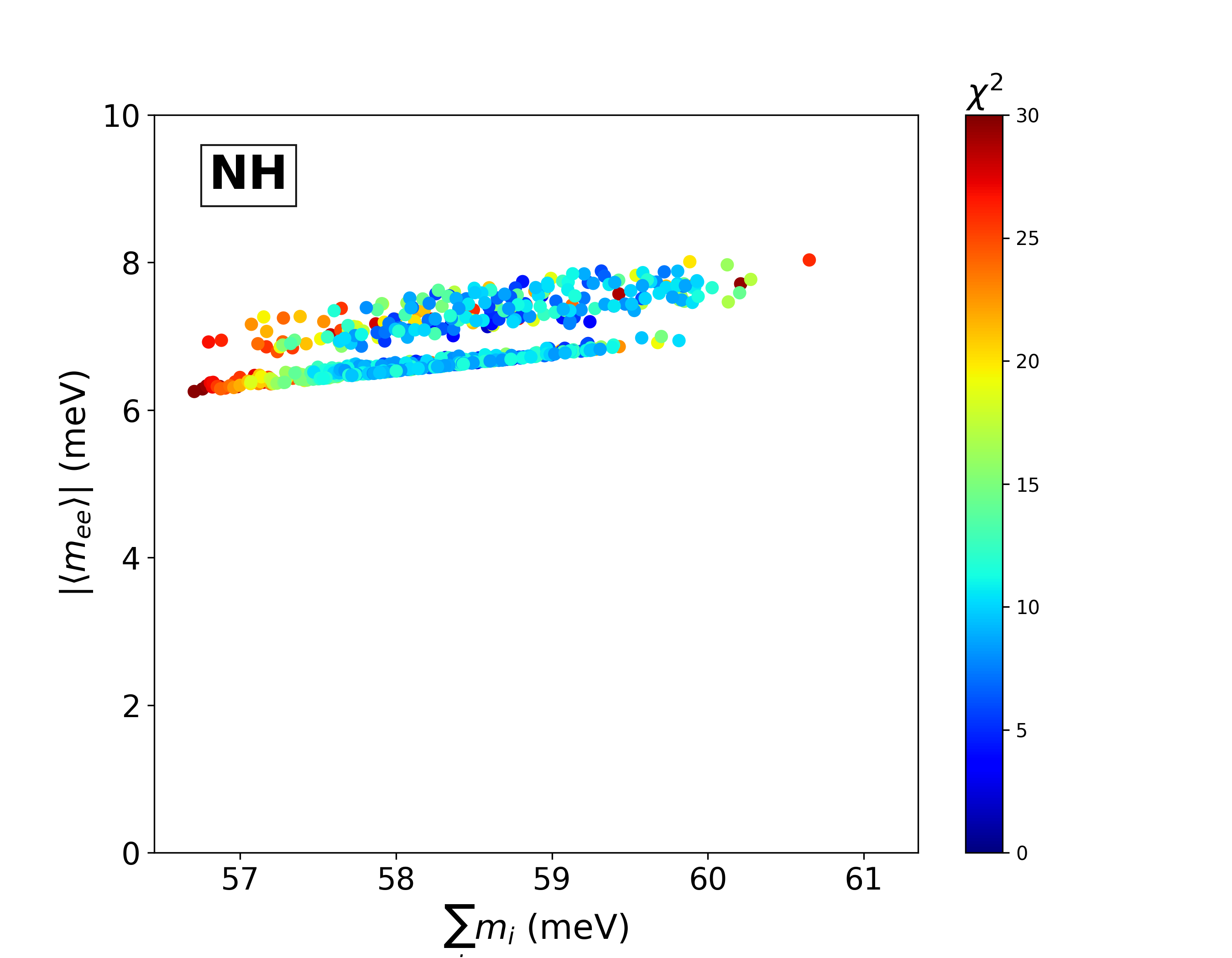}
		\caption{The effective Majorana electron neutrino mass, $\lvert \langle m_{ee} \rangle \rvert$  as a function of the sum of light neutrino masses, $\sum_i m_i$.}
		\label{fig:4}
	\end{center}
\end{figure}

The effective Majorana neutrino mass $\lvert \langle m_{ee} \rangle \rvert$ that characterizes the process of neutrino-less double beta decay ($0\nu \beta\beta$) is given by

\begin{equation}
\label{eq:3.6}
	\lvert \langle m_{ee} \rangle \rvert = \lvert \sum_i m_i U_{ei}^2 \rvert = \lvert c_{12}^2 c_{13}^2 m_1 e^{i \alpha_M} + s_{12}^2 c_{13}^2 m_2 e^{i \beta_M} + s_{13}^2 e^{-2i\delta_{CP}} \rvert = \lvert \left(m_\nu\right)_{11} \rvert
\end{equation}

From equation \ref{eq:3.6} we can see that the effective Majorana neutrino mass depends on the Majorana phases as well as the Dirac CP phase and can be given as the (1, 1) element of the neutrino mass matrix of equation (\ref{eq:2.5}). Using the parameter space of Figure \ref{fig:1} we evaluated the $\lvert \langle m_{ee} \rangle \rvert$ and the results are represented in Figure \ref{fig:4}. The predicted values of $\lvert \langle m_{ee} \rangle \rvert$ lie between ($6.25-8$) meV and it is well below the sensitivity reach of $0\nu \beta\beta$ experiments. 

\section{Conclusion}
\label{sec:4}
This paper examined the minimal form of the inverse seesaw model ISS(2, 2) with $S_4$ flavour symmetry. The $S_4$ flavour symmetry aids in determining the texture of the mass matrices and eventually describing the mixing pattern in the leptonic sector. We performed a test and studied how well the model describes the experimental data using $\chi^2$ analysis. We found that the model describes the experimental neutrino data for NH of neutrino masses with the best-fit value at $\chi^2_{min}\approx 0.24$. The model, however, rules out the case of IH of neutrino masses, with $\chi^2_{min}>100$. The prediction of the Dirac CP phase at the best-fit point is $\delta_{CP}\approx 370.087^\circ$ which can be tested in future precision experiments. Prediction of the model on effective Majorana neutrino mass is also made. The points in the parameter space that satisfy $\chi^2 \leq 30$ have been considered the allowed region for the model parameters. Using this allowed region of the model parameters we evaluate the effective Majorana neutrino mass and found that the obtained values are very small to be tested in future experiments. Experiments such as T2K and NO$\nu$A can resolve the octant of the mixing angle $\theta_{23}$ and give a precision measurement on Dirac CP-violating phase $\delta_{CP}$, which will help us validate our model. Further, the constrained parameter space obtained from our model may be used to study low-scale leptogenesis and is left for future work.
\acknowledgments

BT acknowledges the DST, Government of India for INSPIRE Fellowship
vide Grant no. DST/INSPIRE/2018/IF180588. The research of NKF
is funded by DST-SERB, India under Grant no. EMR/2015/001683.

\bibliographystyle{JHEP}
\bibliography{references.bib}
\end{document}